\documentclass[useAMS,usenatbib,onecolumn]{mn2e}

\usepackage{psfig}
\usepackage{txfonts}
\newcommand{\rd}{{\rm d}}

\title[SPH with flux-limited diffusion]{A faster algorithm for smoothed particle hydrodynamics with radiative transfer in the flux-limited diffusion approximation}

\author[S. C. Whitehouse, M. R. Bate and J. J. Monaghan]{Stuart C. Whitehouse$^1$, Matthew R.
Bate$^1$ and Joe J. Monaghan$^2$\thanks{E-mail: scw@astro.ex.ac.uk,
mbate@astro.ex.ac.uk, joe.monaghan@sci.monash.edu.au}\\ 1. School of Physics, University of Exeter, Stocker
Road, Exeter EX4 4QL\\ 2. School of Mathematical Sciences, Monash University, Clayton 3800, Australia }

\date{Accepted for publication in MNRAS}

\begin{document}
\maketitle

\begin{abstract}

We describe a new, faster implicit algorithm for solving the radiation
hydrodynamics  equations in the flux-limited diffusion approximation for
smoothed particle hydrodynamics.  This improves on the method elucidated in
Whitehouse \& Bate by using a Gauss-Seidel iterative method rather than iterating
over the exchange of energy between pairs of particles. The new algorithm is
typically many thousands of times faster than the old one, which will enable
more complex problems to be solved. The new algorithm is tested using the same
tests performed by Turner \& Stone for {\sc ZEUS-2D}, and repeated by Whitehouse
\& Bate.

\end{abstract}

\begin{keywords}
  hydrodynamics -- methods: numerical -- radiative transfer.
\end{keywords}

\section{Introduction}

Smoothed particle hydrodynamics (SPH) is a Lagrangian method introduced by 
\citet{L1977} and \citet{GM1977} that is frequently used for simulating 
astrophysical fluid problems, for example for star and galaxy formation 
and supernovae. However, it has also been used outside of astrophysics, 
for example for the modelling of tsunami and volcanoes 
\citep[see][for a review]{M1992}.

SPH is normally used to model hydrodynamics, with the inclusion of self-gravity
in astrophysical contexts. The smoothing length, which may be variable both in
space and in time, enables SPH to naturally adapt its resolution to reflect the
local density distribution. However, in many astrophysical situations 
radiation transport is also an important element. 
Various attempts have been made to include radiation transport into SPH.
\citet{L1977} made the first foray into this field, using the diffusion
approximation to examine the fission of protostars. \citet{B1985,B1986}
also used radiative transfer in the diffusion approximation, modelling
stars with entropy gradients.
\citet{OW2003} combined SPH with a Monte-Carlo radiation transport method.
\citet{Viau2001} and \citet{BCV2004} presented an implicit scheme for the 
diffusion approximation.
Finally, \citet{WB2004} developed an implicit scheme for two-temperature
(gas and radiation) radiative transfer in the flux-limited diffusion 
approximation.
%However the method described by \citet{WB2004} was fairly slow,
%and although it accurately performed the tests described therein, some tests
%took an extremely long time.

%\citet{BCV2004} included various thermodynamics to
%simulate cylindrical cloud collapse. \citet{WB2004} used a flux-limiter (c.f.
%\citet{TS2001}) to limit the speed of propagation of radiation in the optically
%thin regime to the speed of light, and included separate temperatures for the
%gas and the radiation.

In this paper we describe a significant improvement to the method of
\citet{WB2004}, namely the implementation of an implicit algorithm which 
is many thousands of times faster, but still models the same physics.  
Section \ref{sec:method}  begins by summarising the origins of the flux-limited
diffusion approximation, and continues into a brief overview of the explicit
SPH equations derived by \citet{WB2004}, before describing in detail the new 
method.  Section \ref{sec:testcalc} describes the tests we performed 
to validate this code. These test results should be compared with 
those of \citet{TS2001} and \citet{WB2004}.

\section{Method}
\label{sec:method}
In a frame co-moving with the fluid, and assuming local thermal equilibrium
(LTE), the equations governing the time-evolution of radiation hydrodynamics
(RHD) are
\begin{equation}
\label{rhd1}
\frac{D\rho}{Dt} + \rho\mbox{\boldmath $\nabla\cdot v$} = 0~,
\end{equation}
\begin{equation}
\label{rhd2}
\rho \frac{D\mbox{\boldmath $v$}}{Dt} = -\nabla p + \frac{\mbox{${\chi_{}}_{\rm \scriptscriptstyle F}\rho$}}{c} \mbox{\boldmath $F$}~,
\end{equation}
\begin{equation}
\label{rhd3}
\rho \frac{D}{Dt}\left( \frac{E}{\rho}\right) = -\mbox{\boldmath $\nabla\cdot F$} - \mbox{\boldmath $\nabla v${\bf :P}} + 4\pi \kappa_{\rm \scriptscriptstyle P} \rho B - c \kappa_{\rm \scriptscriptstyle E} \rho E~,
\end{equation}
\begin{equation}
\label{rhd4}
\rho \frac{D}{Dt}\left( \frac{e}{\rho}\right) = -p \mbox{\boldmath $\nabla\cdot v$} - 4\pi \kappa_{\rm \scriptscriptstyle P} \rho B + c \kappa_{\rm \scriptscriptstyle E} \rho E~,
\end{equation}
\begin{equation}
\label{rhd5}
\frac{\rho}{c^2} \frac{D}{Dt}\left( \frac{\mbox{\boldmath $F$}}{\rho}\right) = -\mbox{\boldmath $\nabla\cdot${\bf P}} - \frac{\mbox{${\chi_{}}_{\rm \scriptscriptstyle F}\rho $}}{c} \mbox{\boldmath $F$}~
\end{equation}
 \citep{MM1984,TS2001}.  In these equations, 
$D/Dt \equiv \partial/\partial t + \mbox{\boldmath $v \cdot \nabla$}$  is 
the convective derivative.  The symbols $\rho$, $e$, {\boldmath $v$} 
and $p$ represent the material mass density, energy density, 
velocity, and scalar isotropic pressure respectively.  The total 
frequency-integrated radiation energy density, momentum density (flux)
and pressure tensor are represented by $E$, {\boldmath $F$}, and {\bf P},
respectively. 

A detailed explanation of the flux-limited diffusion approximation to the  above
equations is given by \citet{TS2001}.   Here we simply summarise the main
points.   The assumption of LTE allows the rate of emission of radiation from
the matter in equations \ref{rhd3} and \ref{rhd4} to be written as the Planck
function, $B$.  Equations \ref{rhd2} to \ref{rhd5} have been integrated over
frequency, leading to the flux mean total opacity ${\chi_{}}_{\rm
\scriptscriptstyle F}$, and the Planck mean and energy mean absorption
opacities, $\kappa_{\rm \scriptscriptstyle P}$ and  $\kappa_{\rm
\scriptscriptstyle E}$.  In this paper, the opacities are assumed to be
independent of frequency so that  $\kappa_{\rm \scriptscriptstyle P}=\kappa_{\rm
\scriptscriptstyle E}$ and the subscripts may be omitted.  The total opacity,
$\chi$, is the sum of components due to the absorption $\kappa$ and the
scattering $\sigma$.

%Note that the above opacities have dimensions of length squared over mass 
%(i.e.\ cm$^2$/g) whereas \citet{TS2001} define their opacities to
%have dimensions of inverse length (i.e.\ their opacities are equal to
%ours multiplied by the density).

The equations of RHD may be closed by an equation of state, specifying the gas
pressure, the addition of constitutive relations for the Planck function  and
opacities, and an assumption about the relationship between the angular moments
of the radiation field. 

In this paper, we use an ideal equation of state for the gas pressure $p =
(\gamma -1)u\rho$, where $u=e/\rho$ is the  specific energy of the gas.  Thus,
the temperature of the gas is  $T_{\rm g}=(\gamma -1)\mu u/R_{\rm g}=u/c_{\rm
v}$, where $\mu$ is the  dimensionless  mean particle mass, $R_{\rm g}$ is the
gas constant and $c_{\rm v}$ is the specific heat capacity of the gas.  The
Planck function is given by $B=(\sigma_{\rm \scriptscriptstyle B}/\pi)T_{\rm
g}^4$,  where $\sigma_{\rm \scriptscriptstyle B}$ is the Stefan-Boltzmann
constant. The radiation energy density also has an associated temperature
$T_{\rm r}$  from the equation $E=4 \sigma_{\rm \scriptscriptstyle B} T_{\rm
r}^4/c$.

For an isotropic radiation field ${\bf \rm P} = \frac{1}{3} E$.  The
Eddington approximation assumes this relation holds everywhere and implies
that, in a steady state, equation \ref{rhd5} becomes
\begin{equation}
\label{eddington}
\mbox{\boldmath $F$} = -\frac{c}{3\chi\rho} \nabla E.
\end{equation}
This expression gives the correct flux in optically thick regions, where
$\chi\rho$ is large. However in optically thin regions where 
$\chi\rho \rightarrow 0$ the flux tends to infinity whereas in reality 
$|\mbox{\boldmath $F$}| \le cE$.  Flux-limited diffusion solves this problem
by limiting the flux in optically thin environments to always obey this
inequality.  \citet{LP1981} wrote the radiation flux
in the form of Fick's law of diffusion as
\begin{equation}
\label{fld1}
\mbox{\boldmath $F$} = -D \nabla E,
\end{equation}
with a diffusion constant given by
\begin{equation}
\label{fld2}
D = \frac{c\lambda}{\chi\rho}.
\end{equation}
The dimensionless function $\lambda(E)$ is called the flux limiter.  The 
radiation pressure tensor may then be written in terms of the radiation
energy density as
\begin{equation}
\label{fld3}
\mbox{\rm \bf P} = \mbox{ \rm \bf f} E,
\end{equation}
where the components of the Eddington tensor, {\bf f}, are given by
\begin{equation}
\label{fld4}
\mbox{\rm \bf f} = \frac{1}{2}(1-f)\mbox{\bf I} + \frac{1}{2}(3f-1)\mbox{\boldmath $\hat{n}\hat{n}$},
\end{equation}
where $\mbox{\boldmath $\hat{n}$}=\nabla E/|\nabla E|$ is the unit vector
in the direction of the radiation energy density gradient and the dimensionless
scalar function $f(E)$ is called the Eddington factor.  The flux limiter
and the Eddington factor are related by
\begin{equation}
\label{fld5}
f = \lambda + \lambda^2 R^2,
\end{equation}
where $R$ is the dimensionless quantity $R = |\nabla E|/(\chi\rho E)$.

Equations \ref{fld1} to \ref{fld5} close the equations of RHD, eliminating
the need to solve equation \ref{rhd5}.  However, we must still choose
an expression for the flux limiter, $\lambda$.  In this paper, we choose
\citet{LP1981}'s flux limiter 
\begin{equation}
\lambda(R) = \frac{2+R}{6 + 3R + R^2},
\end{equation}
to allow comparison of our results with those of \citet{TS2001} and
\citet{WB2004}.  

\subsection{The explicit method}

In \citet{WB2004}, we described a method by which equations \ref{rhd2} 
to \ref{rhd4} can be written in SPH formalism. Equation
\ref{rhd1} does not need to be solved directly since the density of each 
particle is calculated using the standard SPH summation over the particle
and its neighbours.  We define the 
specific energy of the gas to be $u = e /\rho$, and that of the 
radiation to be $\xi = E / \rho$. The explicit equations in one 
dimension are then
\begin{equation}
{D \mbox{\boldmath $v_{i}$} \over D t}  = - \sum_{j=1}^{N} m_{j} \left( {p_{i} \over \rho_{i}^2}
+ { p_{j} \over \rho_{j}^2} + \Pi_{ij} \right) \nabla W(r_{ij},h_{ij}) - 
\frac{\lambda_{i}}{\rho_{i}} \sum_{j=1}^N m_j \xi_{j} \nabla W(r_{ij},h_{ij})~,
\end{equation}

\begin{equation}
\label{eqn:SPHRTE}
{D \xi_i \over D t} =  \sum_{j=1}^N { m_j \over \rho_i \rho_j} c  \left[ {4 {\lambda_i \over
\kappa_{i} \rho_i} {\lambda_j \over \kappa_{j} \rho_j} \over \left( {\lambda_i
\over \kappa_{i} \rho_i} +{\lambda_j \over \kappa_{j} \rho_j} \right) } \right]
\left( \rho_i \xi_i - \rho_j \xi_j \right)  {\nabla W_{ij} \over r_{ij}}  
- \left( \mbox{\boldmath $\nabla \cdot v$} \right)_i f_i
\xi_i - a c
\kappa_{i} \left({\rho_i \xi_i \over a} - \left( {u_i \over c_{{\rm v},i} }
\right)^4 \right),
\end{equation}

\begin{equation}
\label{eqn:SPHRTU}
{D u_i \over D t} = \frac{1}{2} \sum_{j=1}^N \left( {p_i \over
\rho_i^2 } + {p_j \over \rho_j^2} + \Pi_{ij} \right) m_j
\mbox{\boldmath $v$}_{ij} \cdot \nabla W_{ij} + a c \kappa_{i}
\left( {\rho_i \xi_i \over a} - \left( {u_i \over c_{{\rm v},i} } 
\right)^4 \right),
\end{equation}
where $a=4 \sigma_{\rm B}/c$, $m_i$ is the mass of SPH particle $i$, 
$\mbox{\boldmath $r$}_{ij}=\mbox{\boldmath $r$}_{i}-\mbox{\boldmath $r$}_{j}$ 
is the difference in positions between particles $i$ and $j$,
$\mbox{\boldmath $v$}_{ij}=\mbox{\boldmath $v$}_{i}-\mbox{\boldmath $v$}_{j}$,
and $W_{ij}=W(r_{ij}, h_{ij})$,  where $W$ is the standard cubic spline 
kernel and the mean smoothing length of particles $i$ and $j$ is 
$h_{ij}=(h_i + h_j)/2$.  The smoothing lengths are defined in the same 
manner as those in \citet{WB2004}, so each particle has approximately 
eight neighbours unless otherwise stated.

We use the standard SPH artificial viscosity
 
\[ \Pi_{ij} =  \left\{ \begin{array}{ll}
         \left( - \alpha_{\rm v} c_{\rm s} \mbox{ $\mu_{ij}$} + \beta_{\rm v} \mbox{ $\mu_{ij}$}^2 \right) / \rho_{ij}
 & \mbox{if  \boldmath $v$} \mbox{$_{ij} \cdot$}\mbox{\boldmath{ $r$}} \mbox{$_{ij} \leq  0 $  , and } \\
        0  & \mbox{if  \boldmath $v$} \mbox{$_{ij} \cdot$}\mbox{\boldmath{ $r$}} \mbox{$_{ij} > 0$ }. \\
\end{array} \right. \] 
where $\mbox{ $\mu_{ij}$} = { h \left( \mbox{\boldmath $v$}_i - \mbox{\boldmath $v$}_j \right) \cdot \left( \mbox{\boldmath $r$}_i - \mbox{\boldmath $r$}_j \right)
/ \left( \left| \mbox{\boldmath $r$}_i - \mbox{\boldmath $r$}_j \right|^2 + \eta^2 \right)}$, with $\eta^2 = 0.01 h^2$ to prevent numerical divergences if particles get too close together. We use $\alpha_{\rm v}=1$ and $\beta_{\rm v}=2$ unless stated otherwise.
In equation \ref{eqn:SPHRTE} the
first term on the right hand side is the diffusion term, the second is
the work done on the radiation field (in one dimension), and
the final term allows energy transfer between the radiation and the
gas. In equation \ref{eqn:SPHRTU} the energy transfer term occurs with
opposite sign, while the remaining term is the symmetric SPH
expression for work and viscous dissipation done on the gas  when the
thermodynamic variable of integration is energy.

We also tested a supercritical shock (see section \ref{sec:supershock}) with an
implicit method derived from the asymmetric variant of the work term in equation
\ref{eqn:SPHRTU}
\begin{equation}
\label{eqn:assym}
{D u_i \over D t} =  \sum_{j=1}^N \left( 
 {p_i \over \rho_i^2} + \frac{1}{2} \Pi_{ij} \right) m_j
\mbox{\boldmath $v$}_{ij} \cdot \nabla W_{ij} + a c \kappa_{i}
\left( {\rho_i \xi_i \over a} - \left( {u_i \over c_{{\rm v},i} } 
\right)^4 \right).
\end{equation}
The two forms of the gas energy equation gave results that differed by
two per cent or less.  All results presented in this paper use the 
symmetric version.

\subsection{Implicit method}

The implicit method for solving the energy equations 
described by \citet{WB2004} calculated the gas work and viscous terms and 
the diffusion term as an interaction between pairs of particles, 
subtracting energy from one particle and adding it to another. 
The radiation pressure term was added to $\xi$, while the interaction 
term between the gas and the radiation was calculated by the solution 
of a quartic equation.  The required timestep ${\rm d}t$ was split into 
$N$ substeps and the particles swept over.  This solution was then compared
to that with $2N$ substeps.  If the fractional error between the two solutions
was not less than a specified tolerance, the number of substeps was doubled
until the required tolerance was reached.

The new formulation uses a Gauss-Seidel method to iterate towards the solution
of the system of equations.  We use a backwards Euler implicit method rather
than the trapezoidal method used in \citet{WB2004}, because the former allows
larger timesteps to be taken.   To advance a time-dependent variable $A$, 
from time $t=n$ to $t=n+1$, the backwards Euler scheme states
\begin{equation}
\label{eqn:backwardsE}
{ A_i^{n+1} } = A_i^n + \rd t \left(\frac{ \rd A_i}{\rd t}\right)^{n+1}.
\end{equation}
For a Gauss-Seidel method involving interactions between particles $i$
and $j$, the new value $A_i^{n+1}$ can be solved for by arranging the
implicit equations into the form
\begin{equation}
\label{eqn:GS}
A_i^{n+1} = \frac{ A_i^n - \rd t \sum_j \sigma_{ij} \left( A_j^{n+1} 
\right) }{ 1 - \rd t \sum_j \sigma_{ij}},
\end{equation}
where $\sigma_{ij}$ contains quantities other than $A$, and $A_j^{n+1}$ 
begins as $A_j^n$ and is updated as soon as new values become available.
This equation is iterated over until convergence is achieved.

The backwards Euler form of equation \ref{eqn:SPHRTE} is given by
\begin{equation}
\label{eqn:GSxi}
\xi^{n+1}_i =  \xi_i^n + \rd t \sum_j \frac{m_j}{\rho_i \rho_j} b c \left( \rho_i
\xi_i^{n+1} - \rho_j \xi_j^{n+1} \right) \frac{\nabla W_{ij}}{r_{ij}} - \rd t
\left( \mbox{\boldmath $\nabla \cdot v$} \right)_i f_i \xi_i^{n+1} - \rd t a c \kappa_i \left[ \frac{ \rho_i
\xi_i^{n+1}}{a} - \left( \frac{ u_i^{n+1}}{c_{{\rm v},i}} \right)^4 \right],
\end{equation}
where 
\begin{equation}
b=\left[ {4 {\lambda_{i} \over \kappa_{i} \rho_{i}} {\lambda_{j} \over 
\kappa_{j} \rho_{j}} \over \left( {\lambda_{i} \over \kappa_{i} \rho_{i}} +
{\lambda_{j} \over \kappa_{j} \rho_{j}} \right) } \right]
\end{equation}
for brevity, and of equation \ref{eqn:SPHRTU} as
\begin{equation}
\label{eqn:GSu}
u^{n+1}_i = u_i^n + \rd t \sum_j \frac{1}{2} m_j \mbox{\boldmath $v_{ij}$} \cdot \nabla W_{ij} \left(
\frac{ u_i^{n+1} \left( \gamma - 1 \right) }{\rho_i} + \frac{ u_j^{n+1} 
\left( \gamma - 1 \right) }{\rho_j} + \Pi_{ij} \right) +  \rd t a c \kappa_i \left[ \frac{ \rho_i
\xi_i^{n+1}}{a} - \left( \frac{ u_i^{n+1}}{c_{{\rm v},i}} \right)^4 \right],
\end{equation}
substituting in $(\gamma -1)u\rho$ for $p$ using our equation of state.  These
equations can be rearranged into the form of equation \ref{eqn:GS}
to solve for $\xi^{n+1}_i$ and $u^{n+1}_i$ and Gauss-Seidel iteration 
performed with all other independent variables fixed.

We investigated two different approaches to solving these equations. The first 
method was to iterate over the Gauss-Seidel form of equations \ref{eqn:GSxi} 
and \ref{eqn:GSu} separately but within the same iterations.  This resulted 
in implicit integration that was many times faster than that of \citet{WB2004} 
in optically thin regimes, but in optically thick regions the performance of 
the code was similar to that of \citet{WB2004}.  The method also failed to
converge for large timesteps when the energy transfer term (the last terms in
equations \ref{eqn:GSxi} and \ref{eqn:GSu}) became large (due to high 
$\kappa$ or large temperature differences between the gas and the radiation).

By far the most effective method is to solve equations \ref{eqn:GSxi} and
\ref{eqn:GSu} simultaneously for $u_i^{n+1}$, and to perform Gauss-Seidel 
iteration on the resulting expression.  The resulting value of $u_i^{n+1}$ 
is then substituted into equation \ref{eqn:GSxi} to obtain $\xi_i^{n+1}$
during the same iteration.  To simplify the subsequent equations we define 
the following quantities:
\begin{eqnarray} 
\beta = & \displaystyle \rd t ~c~ \kappa_i ~\rho_i ,\nonumber \\
\Gamma = & \displaystyle a~ c~ \kappa_i / c_{{\rm v}, i}^4, \nonumber \\ 
D_{{\rm d},i} = & \displaystyle \sum_j \frac{m_j}{\rho_j} c~b \frac{\nabla
W_{ij}}{r_{ij}}, \nonumber \\
%Note definition includes minus sign! 
D_{{\rm n},i} = & \displaystyle - \sum_j \frac{m_j}{\rho_i~\rho_j} c~b \frac{\nabla
W_{ij}}{r_{ij}} \rho_j \xi_j^{n+1}, \nonumber \\  
%Note definition includes the half!
P_{{\rm d},i} = & \displaystyle \sum_j \frac{1}{2} m_j \mbox{\boldmath $v_{ij}$} \cdot \nabla
W_{ij} \frac{ \left( \gamma - 1 \right)}{\rho_i}, \nonumber \\ 
%Note definition includes the half! 
P_{{\rm n},i} = & \displaystyle \sum_j \frac{1}{2} m_j \mbox{\boldmath $v_{ij}$} \cdot
\nabla W_{ij} \left[ \frac{ \left( \gamma - 1 \right) u_j^{n+1} }{\rho_j} +
\Pi_{ij} \right], \nonumber \\ 
R_{{\rm p},i} = & \displaystyle \left( \mbox{\boldmath $\nabla \cdot v$} \right)_i f_i, \nonumber \\
 \chi = &  \rd t D_{{\rm d},i} -  \rd t
R_{{\rm p},i}. \nonumber 
\end{eqnarray}

%\begin{equation}
%\beta = \rd t c \kappa_i \rho_i ,
%\end{equation}
%\begin{equation}
%\Gamma = a c \kappa_i / c_{{\rm v}, i}^4,
%\end{equation}
%\begin{equation}
%D_{{\rm d},i} = \sum_j \frac{m_j}{\rho_j} c b \frac{\nabla W_{ij}}{r_{ij}}
%\end{equation}
%\begin{equation}
%%Note definition includes minus sign!
%D_{{\rm n},i} = - \sum_j \frac{m_j}{\rho_i \rho_j} c b \frac{\nabla W_{ij}}{r_{ij}} \rho_j
%\xi_j^{n+1},
%\end{equation}
%\begin{equation}
%%Note definition includes the half!
%P_{{\rm d},i} = \sum_j \frac{1}{2} m_j v_{ij} \nabla W_{ij} \frac{ \left( \gamma - 1
%\right)}{\rho_i},
%\end{equation} 
%\begin{equation}
%%Note definition includes the half!
%P_{{\rm n},i} = \sum_j \frac{1}{2} m_j v_{ij} \nabla W_{ij} \left[ \frac{ \left( \gamma - 1
%\right) u_j^{n+1} }{\rho_j} + \Pi_{ij} \right],
%\end{equation}
%\begin{equation}
%R_{{\rm p},i} = \nabla v_i f_i,
%\end{equation}
%and
%\begin{equation}
%\chi = \frac{1}{2} \rd t D_{{\rm d},i} - \frac{1}{2} \rd t R_{{\rm p},i} - \beta.
%\end{equation}

Using these new variables we can solve equation \ref{eqn:GSu} for $\xi_i^{n+1}$
\begin{equation}
\label{eqn:xisubst}
\xi_i^{n+1} = \frac{1}{\beta} \left( u_i^{n+1} - u_i^n - \rd t~ P_{{\rm n},i} -
\rd t ~P_{{\rm d},i} u_i^{n+1} + \rd t ~\Gamma \left[ u_i^{n+1} \right]^4 \right).
\end{equation}
The right hand side of equation \ref{eqn:xisubst} then replaces $\xi_i^{n+1}$ 
in equation \ref{eqn:GSxi}, forming a
quartic equation in $u_i^{n+1}$. If the quartic equation is cast in the form $
a_4 x^4 + a_3 x^3 + a_2 x^2 +a_1 x + a_0 = 0$ then the co-efficients are given
by:
\begin{eqnarray}
a_4 = & \displaystyle \Gamma \rd t ~\left( \chi - 1 \right) \nonumber \\
a_3 = & \displaystyle 0 \nonumber \\
a_2 = & \displaystyle 0 \nonumber \\
a_1 = & \displaystyle \left( \chi - \beta - 1 \right) \left( 1 - \rd t~ P_{{\rm d},i} 
\right) \nonumber \\
a_0 = & \displaystyle \beta \xi^n_i + \left( \chi - \beta - 1 \right) \left( - u_i^n -
\rd t~ P_{{\rm n},i} \right) + \rd t ~D_{{\rm n},i} \beta \nonumber 
\end{eqnarray}
Solving this quartic equation yields a value for $u_i^{n+1}$, which may then
be substituted into
\begin{equation}
\label{eqn:xiinredef}
\xi_i^{n+1} = \frac{ \left( \xi_i^n + \rd t~ D_{{\rm n},i} + \rd t ~\Gamma \left[ u_i^{n+1}
\right]^4  \right) }{ 1 - \chi + \beta},
\end{equation}
using the quantities defined above \citep[for the analytic solution of a
quartic equation, see Appendix A of][]{WB2004}. These solutions for
$\xi_i^{n+1}$ and $u_i^{n+1}$ are iterated until
they converge.

\subsection{Prediction of position, density and smoothing length}
\label{sec:prediction}
We found that the accuracy when taking large implicit timesteps could be
improved by predicting forward many of the quantities on the right-hand 
sides of equations \ref{eqn:GSxi} and \ref{eqn:GSu} to time $t=n+1$.  
The quantities $x$, $\rho$, and $h$ can be predicted forwards as
\begin{eqnarray}
x_i^{n+1} =& x_i^n + \rd t\; \mbox{\boldmath $v^n_i$} \nonumber \\
\rho_i^{n+1} =& \rho_i^n - \rd t\; \rho_i^n \left( \mbox{\boldmath $\nabla \cdot v$} \right)_i^n  \nonumber \\
h_i^{n+1} =& h_i^n +  \rd t\; h_i^n \left( \mbox{\boldmath $\nabla \cdot v$} \right)_i^n.
\end{eqnarray}
The improvement in accuracy was especially apparent in 
the case of the supercritical shock (see section \ref{sec:supershock}).

\subsection{Convergence criteria}

We define convergence as being when the values of $u_i^{n+1}$ and $\xi_i^{n+1}$
obtained from the $m$-th iteration satisfy equations \ref{eqn:GSxi} and 
\ref{eqn:GSu} to a given tolerance (with all occurrences of $u_i^{n+1}$ and $\xi_i^{n+1}$ on the right-hand sides of these equations being evaluated from iteration $m-1$).  Thus, for example, we iterate until the fractional
errors in $\xi$ given by
\begin{equation}
\frac{\xi^{n+1,m}_i - \left( \xi_i^n + \rd t \sum_j \frac{m_j}{\rho_i \rho_j} b c
\left( \rho_i \xi_i^{n+1,m-1} - \rho_j \xi_j^{n+1} \right) \frac{\nabla
W_{ij}}{r_{ij}} - \rd t \left( \mbox{\boldmath $\nabla \cdot v$} \right)_i f_i \xi_i^{n+1,m-1} - \rd t~ a ~c~
\kappa_i \left[ \frac{ \rho_i \xi_i^{n+1,m-1}}{a} - \left( \frac{ u_i^{n+1,m-1}}{c_{{\rm
v},i}} \right)^4 \right] \right)}{\xi^{n+1,m}_i},
\end{equation}
are within a certain tolerance, for which we have typically used $10^{-3}$.

In the event that the method fails to converge, we split the timestep into two
halves, and begin the iterations again, using the result of the first half 
timestep as the input to the second.  If either fails, we split the timesteps
by another factor of two and continue this way until the system
converges, or it reaches some excessively small fraction of the original 
timestep, at which stage it is no longer computationally efficient to continue
the calculation.

\subsection{Timestep criteria}

The integration of the hydrodynamic variables requires that the
timesteps obey the Courant condition for the hydrodynamic processes.
The usual hydrodynamical
SPH timestep criteria are
\begin{equation}
\label{tshydro1}
\rd t_{{\rm Courant},i} = { \zeta h_{i} \over c_{\rm s} + h_{i} \left| \nabla \cdot \mbox{\boldmath $v$} \right|_{i} + 1.2 
\left( \alpha_{\rm v} c_{\rm s} + \beta_{\rm v} h_{i} \left| \nabla \cdot \mbox{\boldmath $v$} \right|_{i} \right) },
\end{equation}
and 
\begin{equation}
\label{tshydro2}
\rd t_{{\rm force},i} = \zeta \sqrt{{h_{i} \over \left| \mbox{\boldmath $a$}_{i}  \right|}},
\end{equation}
where we use a Courant number of $\zeta=0.3$, unless otherwise noted, and 
\mbox{\boldmath $a$}$_i$ is the 
acceleration of particle $i$. The lesser of these two quantities gives the
hydrodynamical timestep.

There is also an explicit timestep associated with the radiation 
hydrodynamics, described in detail in \citet{WB2004}. This timestep is 
typically much smaller than the hydrodynamical timestep, and the new implicit
method enables us to forgo the use of smaller timesteps in favour of the 
large hydrodynamical timestep.

\section{Test calculations}
\label{sec:testcalc}

We have once again duplicated the tests done by \citet{TS2001}, as we did in
\citet{WB2004}, this time however using the new code. This code achieves 
the same or better accuracy than the code described in \citet{WB2004}.
However, in the vast majority of cases (especially those involving
moving fluids) the new code is significantly faster 
than the old one.

\subsection{Heating and cooling terms}
\label{sec:heatcool}

We tested the interaction between the radiation and the 
gas to check that the temperatures of the gas and the radiation equalise
at the correct rate
when $T_{\rm g} \neq T_{\rm r}$ initially.  A gas in a domain 10 cm long with
100 particles was set up so that there was no velocity, with a density
$\rho = 10^{-7}$ g cm$^{-3}$, opacity $\kappa = 0.4$ cm$^2$ g$^{-1}$, and 
$\gamma=
 \frac{5}{3}$ and  $E = \xi \rho = 10^{12}$ ergs cm$^{-3}$. 
Two tests were carried out, one where the gas heated until it
reached the radiation temperature, and one where it cooled. The first test had
$e=u \rho = 10^2$ ergs cm$^{-3}$, and the second 
$e=u \rho = 10^{10}$ ergs cm$^{-3}$. 
The boundaries of the calculation used reflective ghost particles.

This problem can be approximated in the case where the energy in the radiation is much greater than that in the 
gas by the differential equation
\begin{equation}
\frac{\rd e}{\rd t} = c \kappa E - a c \kappa \left( {e  \over \rho c_{\rm v}} \right)^4,
\end{equation}
and assuming $E$ is constant.

\begin{figure}
\centerline{\psfig{figure=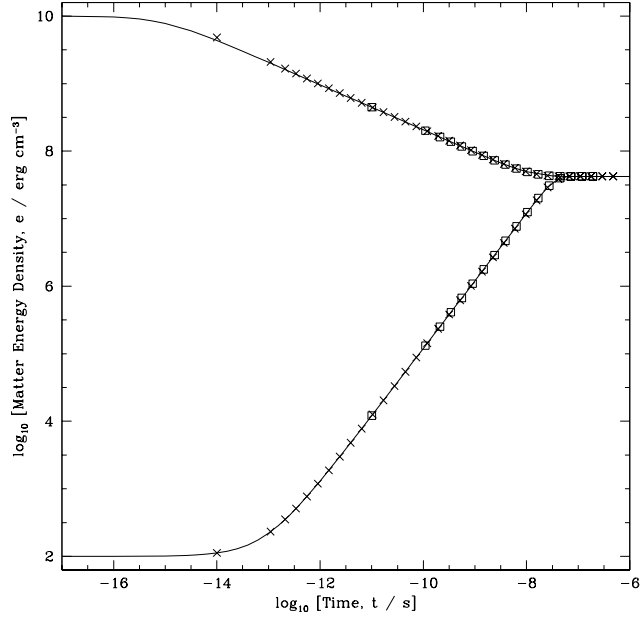,width=9.0truecm}}
\caption{\label{fig:ts51} 
The evolution of the gas energy density $e$ as it equilibriates with a 
radiation energy density $E=10^{12}$ erg cm$^{-3}$.  In the upper case, 
$e=10^{10}$ erg cm$^{-3}$ initially, while in the lower case
$e=10^{2}$ erg cm$^{-3}$. The
solid line is the analytic solution, the crosses are the results of the SPH
code using implicit timesteps of the lesser of $10^{-14}$~s or five percent 
of the 
elapsed time, and the squares with a timestep of the lower of $10^{-11}$~s or
five percent of the elapsed time. The symbols are plotted every ten timesteps.}
\end{figure}

In figure \ref{fig:ts51}, the solid line is this analytic solution, 
plotted both for the cases where $T_{\rm g}$ increases and decreases. The
crosses are the results of the SPH code using an implicit timestep that is 
set to  
the greater of $10^{-14}$ s or five percent of the time elapsed, similar
to the way \citet{WB2004} performed this test. The squares
are similar, but with a timestep being the greater of $10^{-11}$ s or
five percent of the time elapsed.
As can be seen, the match between the analytic solution and the solutions 
given by the SPH code is once again excellent.  On this test, the new method
and that of \citet{WB2004} are of comparable speed and each takes a 
few minutes.

\subsection{Propagating radiation in optically thin media}

\begin{figure}
\centerline{\psfig{figure=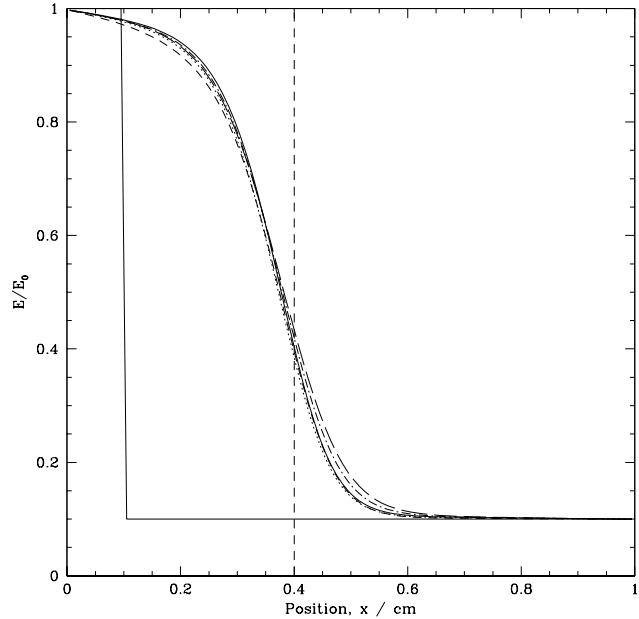,width=9.0truecm}}
\caption{\label{fig:ts55} 
The propagation of a radiation pulse across a uniform medium. The
time is $t=10^{-11}$~s. The vertical dashed line shows the expected position
of the pulse based on the speed of light. The results are almost
independent of the size of the implicit timestep used.  Results are
given for implicit timesteps equal to (solid line),
ten times (dotted line), one hundred times (short-dashed line), and one
thousand times (long-dashed line) the explicit 
timestep.  The dot-dashed line gives the result using a single implicit 
step of $10^{-11}$~s.  
The initial conditions are also shown as a solid line.}
\end{figure}

In the standard diffusion approximation, in optically thin regions, radiation 
propagates at near infinite speed. 
This is unphysical, and the  flux limiter has been
introduced to limit the diffusion of radiation to the speed of light.  To
examine how well our code limits the speed of the radiation, a one 
centimetre long one-dimensional box is filled with 100 equally spaced SPH
particles, with $E= 10^{-2}$ erg cm$^{-3}$ ($\xi = 0.4$ erg g$^{-1}$), $\rho =
0.025$ g cm$^{-3}$, and $ \kappa=0.4$ cm$^2$ g$^{-1}$.  Initially, the radiation
and gas are in thermal equilibrium.

At the start of the simulation, the radiation energy density for the leftmost
ten particles was changed to $E=0.1$ erg cm$^{-3}$($\xi = 4$ erg g$^{-1}$) 
 causing a 
radiation front that was allowed to propagate across the region. 
The ghost particles were reflective except in specific radiation energy
$\xi$,  which was fixed equal to $\xi=4$ erg g$^{-1}$ at the left hand 
boundary and $\xi=0.4$ erg g$^{-1}$ at the right hand boundary.
The implicit code was used
with various timesteps ranging from an explicit timestep to a single
implicit step lasting $10^{-11}$~s. The results are shown in figure 
\ref{fig:ts55}.

As shown by the figure, the radiation pulse propagates at the correct speed,
even using one single large timestep. The front is smoothed out in a
manner similar to the results of \citet{TS2001} and \citet{WB2004}; 
both methods are quite diffusive in this situation.  The new code is slower
than that of \citet{WB2004} for an explicit timestep, but superior for
all longer timesteps.

\subsection{Optically-thick (adiabatic) and optically-thin (isothermal) shocks}

% Old code
%4 days 10 hours viso
%4 dats 7 hours

A shock-tube test, identical to the one in \citet{WB2004}, was set up to
investigate the way the code simulated  optically-thin and optically-thick
regimes and the transition between them. In the limit of high optical depth, the
gas cannot cool because the  radiation is trapped within the gas; thus the shock
is adiabatic.  An  optically-thin shock, on the other hand, is able to
efficiently radiate  away the thermal energy and thus behaves as an isothermal
shock.  In these shock tests, the gas and radiation are highly coupled and,
thus, their  temperatures are equal.

A domain $2 \times 10^{15}$ cm long extending from $x = -1 \times 10^{15}$  to
$1 \times 10^{15} $ cm was set up, with an initial density of $\rho = 10^{-10}$
g cm$^{-3}$, and the temperatures of the  gas and radiation were initially $1500
$~K.  One hundred particles were equally spaced in the domain, with those with
negative $x$ having a velocity equal to the adiabatic ($\gamma = 5/3$) sound
speed $v_0=c_{\rm s}= 3.2\times 10^5$ cm s$^{-1}$, and those with positive $x$
travelling at the same speed in the opposite direction.  The two flows impact at
the origin, and a shock forms.  Opacities of $\kappa = 40, 0.4,  4.0 \times
10^{-3} $  and $4.0 \times 10^{-5}$ cm$^2$ g$^{-1}$ were used to follow the
transition from adiabatic to isothermal behaviour.  Ghost particles were placed
outside the boundaries and maintain the initial energies of their respective 
real particles. The boundaries moved inwards with the same velocity as the
initial velocities of the two streams.  These moving boundaries cause slight
perturbations in the densities of those particles closest to the boundaries,
however this does not affect the solution in the vicinity of the shocks.

\begin{figure}
\centerline{\psfig{figure=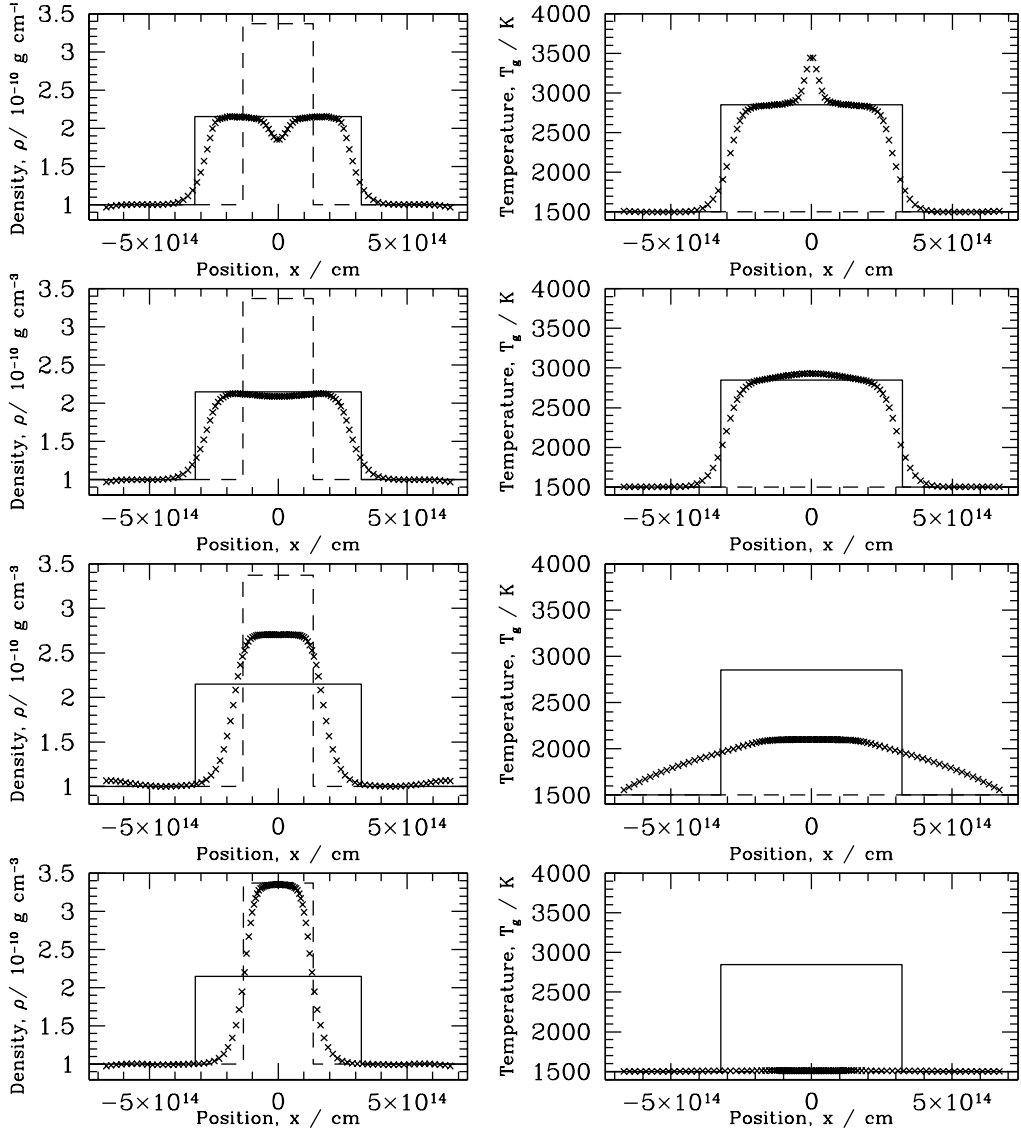,width=15.0truecm}}
\caption{\label{fig:isoadia} 
A set of shocks with differing opacity at time $t=1.0\times 10^9$~s. Density
is on the left, and gas temperature on the right. The crosses are the SPH
results; the solid line gives the analytic solution for an adiabatic shock, and
the dashed line for an isothermal shock. The opacities are (top) $\kappa = 
40$, (upper middle) $\kappa = 0.4 $, (lower middle)  $\kappa = 
4.0 \times 10^{-3}$ and (bottom) $\kappa = 4.0 \times 10^{-5}$ cm$^2$ g$^{-1}$.
As the opacity is decreased, the shocks transition from adiabatic to
isothermal behaviour.}
\end{figure}

%The adiabatic and isothermal limits can be solved analytically 
The optically thick and thin limits can be solved analytically 
(e.g. \citealp{Zeldo}).  The shock speed is given by
\begin{equation}
D = \frac{ ( \gamma_{\rm eff} - 3) + \sqrt{ (\gamma_{\rm eff} +1 )^2 v_0^2 + 16 \gamma_{\rm eff} }}{4},
\end{equation}
where $\gamma_{\rm eff}=1$ 
for the isothermal case, and $\gamma_{\rm eff}=5/3$ for the adiabatic case.  The ratio of the final to the initial density 
is given by 
\begin{equation}
\frac{\rho_1}{\rho_0} = 1 + {v_0 \over D},
\end{equation}
and, for the adiabatic shock, the ratio of the final to the initial temperature
is given by
\begin{equation}
\frac{T_1}{T_0} = \frac{\rho_1}{\rho_0} + \frac{v_0 D \rho_0}{p_0}.
\end{equation}
These analytic solutions are shown by the solid and dashed lines in figure
\ref{fig:isoadia}.  In the figure, the opacity decreases from top to bottom
showing the transition from optically-thick (adiabatic) to optically-thin 
(isothermal) behaviour.  The extremes are in good agreement with their
respective analytic adiabatic and isothermal solutions.  
Note that the
spike in thermal energy near the origin and the corresponding reduction 
in density for the optically-thick case (due to `wall-heating') is softened
by the radiation transport that occurs in the intermediate opacity
calculation with $\kappa = 0.4$.

In comparison to the code described in \citet{WB2004}, these shocks were many
times faster. The timestep was limited by the hydrodynamical criteria only. The
$\kappa = 40$ and $\kappa = 0.4$ shocks ran in less than a minute, compared to
the old code which took thirty-four hours and thirty-five minutes, respectively.
The $\kappa = 4 \times 10^{-3}$ shock took just under  one minute, compared to 
forty-five minutes for the old code.  The $\kappa = 4 \times 10^{-5}$ code took
twenty-three minutes,  compared to the previous time of ten hours.

%Old code:
%Viso: w72/W/2: 10.75 hours
%%%%%%%Iso: /perky/SCRATCH/scw/w69/f/11/cont  2 days 16 + 5 days 20 hours
%Iso: /perky/SCRATCH/scw/w69/f/7/   45 minutes
%Inter: w71/F/2 : 35 minutes
%Adia: w73/Tu/5: 34 hours

%New code:
%Viso: 23 minutes
%Iso: 1 minute
%Inter: < 1 minute
%Adia: < 1 minute

\subsection{Sub- and super-critical shocks}
\label{sec:supershock}
\begin{figure}
\centerline{\psfig{figure=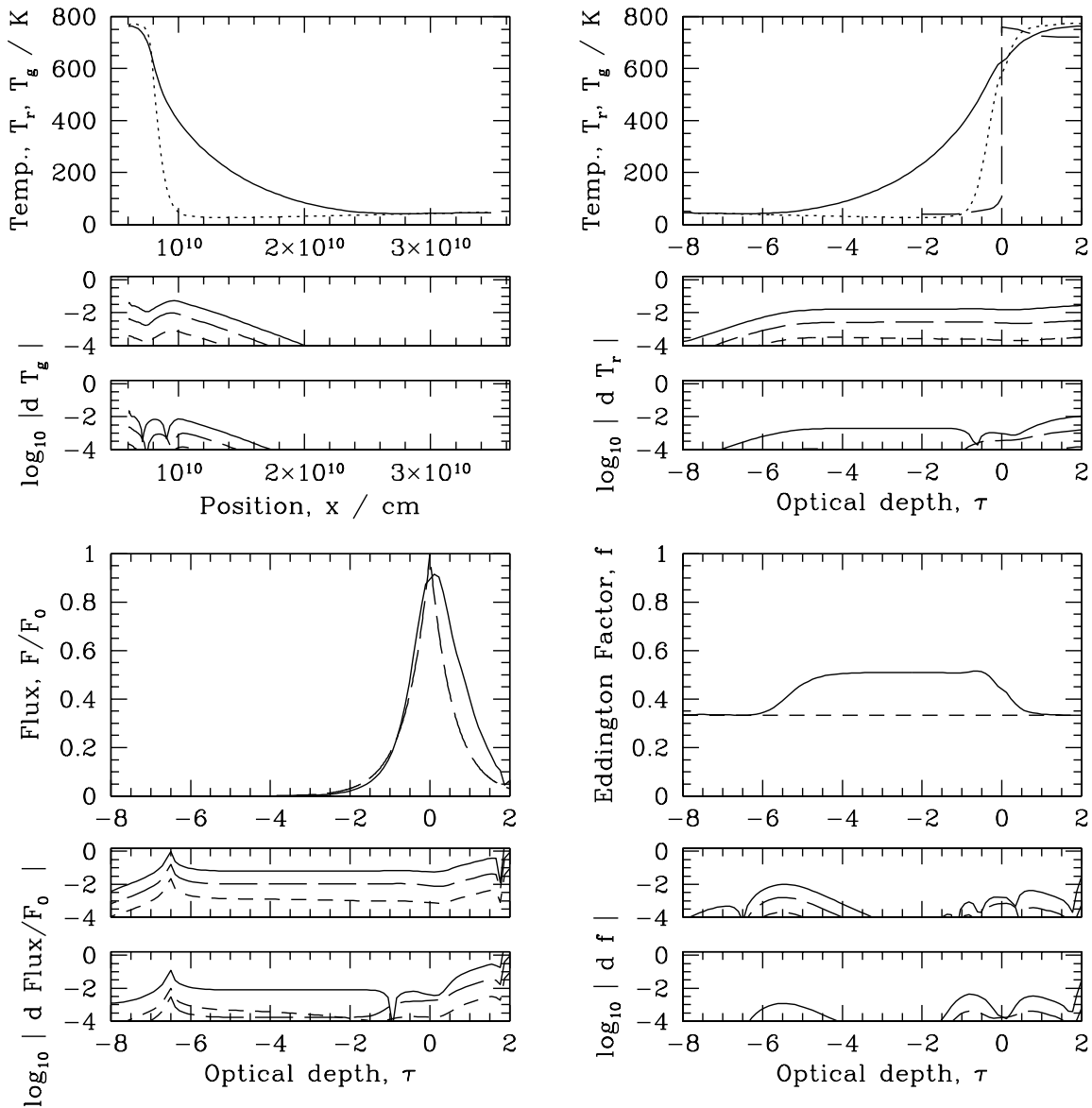,width=16.0truecm}}
\caption{\label{fig:subshock}
The sub-critical shock  
with piston velocity $6 \times 10^5$ cm s$^{-1}$ and
100 particles. The 
large panels show the results using the explicit code. The top panels show 
radiation (solid line) and gas (dotted line) temperatures.  The bottom 
left panel shows the normalised flux and bottom right panel the Eddington 
factor. The long-dashed lines give the analytic solutions for the gas 
temperature and normalised flux.  An Eddington factor of 1/3 is also 
indicated for reference (short-dashed line, lower right panel).  
The subpanels plot the logarithm of the difference between the results 
using the explicit code and 
the implicit code (see the main text). The subpanels show the results with
(lower subpanel) and 
without (upper subpanel) predicting $x$,$h$ and $\rho$ forwards in time.  The implicit code was run with 
timesteps one times (solid line), one tenth of (long-dashed line) and one
hundredth of (short-dashed line) the {\it hydrodynamic} timestep.}
\end{figure}

A supercritical shock occurs when the photons generated by a shock have 
sufficient energy 
to preheat the material upstream. The characteristic temperature profile of
a supercritical shock is where the temperature on either side of the shock
is similar, rather than the downstream temperature being much higher than
that upstream, as occurs in a subcritical shock \citep[see][for more 
details]{Zeldo}. 

The initial conditions of this problem are those of \citet*{SGM1999} and
\citet{TS2001}. A gas with opacity 
$\kappa = 0.4 $ cm$^2$ g$^{-1}$, 
uniform density $\rho=7.78 \times 10^{-10}$ g cm$^{-3}$,  mean molecular
weight $\mu = 0.5$ and $\gamma = \frac{5}{3}$ is set up with $\xi$ and $u$
in equilibrium, with a temperature gradient of $T = 10 + \left( 75 x 
/ 7 \times 10^{10} \right)$ K. Initially the particles are
equally spaced between $x=0$ and $x= 7 \times 10^{10} $ cm for the
supercritical shock, and between $x=0$ and $x= 3.5 \times 10^{10} $ cm for
the subcritical shock. At time $t=0$ a
piston starts to move into the fluid from the left-hand boundary (simulated
by moving the location of the 
boundary). For the subcritical shock the piston velocity is
$v_{\rm p} = 6$ km s$^{-1}$, and for the supercritical shock $v_{\rm p}
 = 16$ km s$^{-1}$, as
per \citet{SGM1999}. The ghost particles are reflective in the frame of
reference of the boundary. Artificial viscosity parameters 
$\alpha_{\rm v} = 2$ and $\beta_{\rm v} = 4$ were used to 
smooth out oscillations.

\begin{figure}
\centerline{\psfig{figure=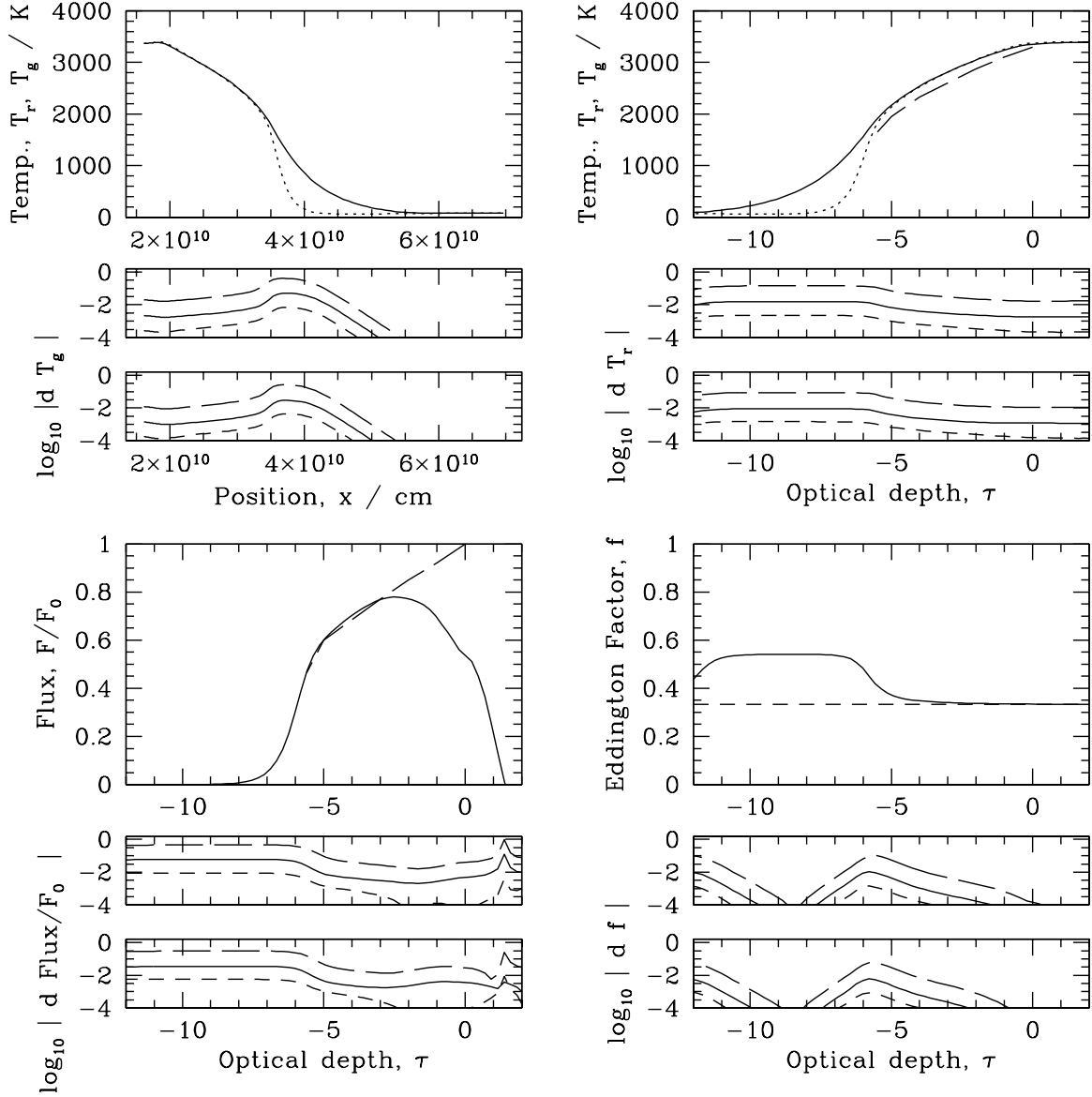,width=16.0truecm}}
\caption{\label{fig:supershock}
The super-critical shock, with piston velocity $1.6 \times 10^6$ cm s$^{-1}$,
100 particles. The subpanels shown the results with (bottom) and without (top) 
predicting $x$,$h$ and $\rho$ forwards in time.  
This shock is strong enough for radiation from the shock 
to preheat the gas upstream.  See figure \ref{fig:subshock} for details 
of the line meanings.}
\end{figure}

\begin{figure}
\centerline{\psfig{figure=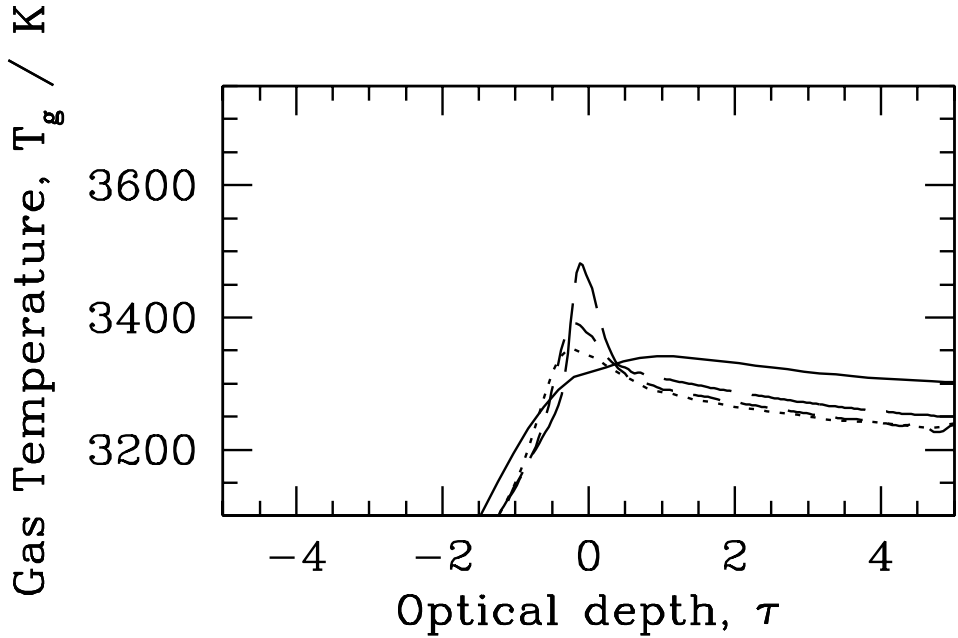,width=16.0truecm}}
\caption{\label{fig:supershockres}
The super-critical shock, with piston velocity $16$ km s$^{-1}$,
showing how changing the resolution affects the spike in gas temperature at
the shock.  All calculations use a hydrodynamical timestep and the implicit code.
The solid line is with 100 particles, the dotted line 
with 200 particles, and the long-dashed with
500 particles. The short-dashed line shows the results with 500 particles
and double the number of neighbours (16 instead of 8).} 
\end{figure}

The results of calculations for a sub-critical shock (piston velocity 
$v_{\rm p} = 6$ km s$^{-1}$) are shown in figure \ref{fig:subshock}.
The 
top left panel for each shows the temperature of the radiation field (solid
line) and the gas (dotted line) against position, and the top right shows the
same quantities against optical depth $\tau$, with $\tau=0$ set at the
shock front (measured from the density distribution). The bottom left panel 
shows normalised flux, and the bottom right the value of the
Eddington factor.  The analytic solutions discussed by \citet{SGM1999} and
\citet{Zeldo} for the temperatures and fluxes of the shocks are
shown with long-dashed lines.
Figures \ref{fig:subshock} and \ref{fig:supershock}
are plotted using the explicit code.  In subpanels beneath the main panels,
we compare the results from the implicit code with the explicit results.
Calculations were performed using the implicit code with timesteps of 
one, one tenth and one hundredth
 the hydrodynamical timestep criteria.  In the subpanels, we plot the differences of the implicit
results with respect to the explicit results.  We divide the difference 
between the implicit and explicit values by the explicit value to obtain a
fractional error and take the logarithm of the absolute value of this 
fraction.  Thus, a difference of $-2$ on the subpanels corresponds to 
an error of 1 percent with respect to the explicit result. 
The subpanels in figure \ref{fig:subshock} shows the log fractional error of
an implicit code with (bottom subpanels) and without (top subpanels)
the prediction of $x$, $h$ and $\rho$ discussed in
section \ref{sec:prediction}.  
Similarly, figure \ref{fig:supershock}  show the
super-critical shock in the same way, excluding (top) and 
including (bottom) the prediction mentioned in section \ref{sec:prediction}.
The prediction of $x$, $h$ and $\rho$ makes the sub-critical shock 
significantly more accurate for hydrodynamical timesteps, whilst having 
a smaller benefit for the super-critical shock. Figure \ref{fig:supershockres}
shows how increasing the resolution of the simulation enables us to resolve
the spike in gas temperature at the shock front in a super-critical shock.

%w70/th/7 and w73/tu/4 
 
The new implicit method is many times faster than the old method. In 
\citet{WB2004}, the super-critical shock with the hydrodynamic timestep 
criteria
took nearly ten days. The new code performed the same calculation on a
comparable CPU in two minutes, making the new algorithm  approximately $10^4$
times faster. A similar improvement in performance can be seen with the sub-critical 
shock -- the hydrodynamic timestep run took over twenty-three days for
\citet{WB2004}, while the present method ran in four minutes, yielding an
increase again of $\approx 10^4$ times.

\subsection{Radiation-dominated shock}

%Old code ~44 hours

In material of high optical depth the radiation generated in a shock cannot
diffuse away at a high rate, and so the radiation becomes confined in a thin
region adjacent to the shock.  \citet{TS2001}  performed a calculation that
tests whether the shock thickness is what one  would expect in these
circumstances.  An extremely high Mach number shock (Mach number of 658)  is set
up, with the gas on the left having an initial density of  $\rho = 0.01$ g
cm$^{-3}$, opacity $\kappa=0.4$ cm$^2$
g$^{-1}$, temperature $T_{\rm r} = T_{\rm g} =  10^4$ K, and speed $10^9$ cm
s$^{-1}$. The gas on the right has density  $\rho = 0.0685847$ g cm$^{-3}$,
opacity $\kappa=0.4$ cm$^2$ g$^{-1}$,  temperature $T_{\rm r} = T_{\rm g} =
4.239 \times 10^7$ K, and speed $1.458 \times 10^8$ cm s$^{-1}$. 
The density contrast is set initially by using different mass particles, however
as the simulation evolves particles from the left enter the shock and by the
time the figure is plotted all the particles shown have the same mass.
The locations
of the boundaries  move with the same speed as their respective particles, and
the properties  of the ghost particles outside these boundaries are fixed at
their initial values. We use hydrodynamical timestep with a Courant number of
$\zeta = 0.03$.   1500 particles are equally spaced over a domain extending from
$x=-6 \times 10^5$ cm to $x=1.5 \times 10^5$ cm  initially, with the
discontinuity at $x = 0.5 \times 10^5$ cm.   The location of the shock should be
fixed in this frame, although  individual particles flow through the shock.

After a period where a transient feature forms at the shock front 
and drifts downstream with the flow, a stable shock is established. Its 
thickness is expected to be roughly equal to the distance 
$l = { c \lambda / \kappa \rho u_1}$, 
where $u_1$ is the speed of material flowing into the shock front.

\begin{figure}
\centerline{\psfig{figure=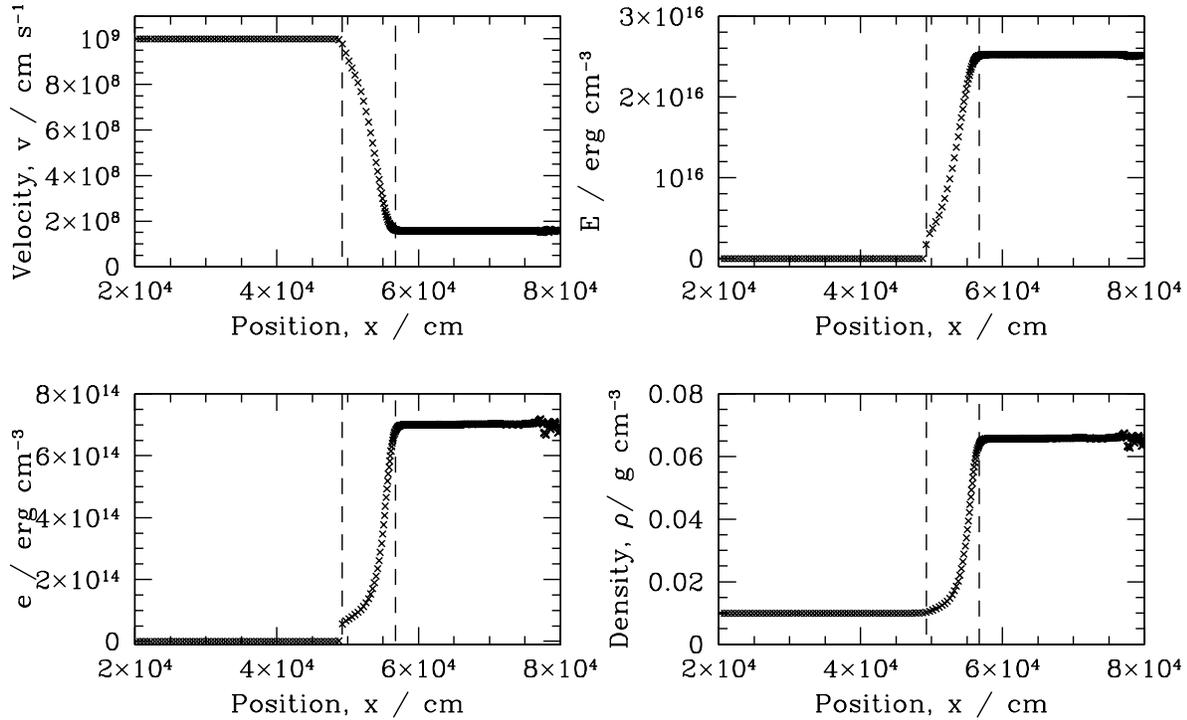,width=16.0truecm}}
\caption{\label{fig:radthick} The radiation-dominated shock at time 
$t= 5 \times 10^{-4}$~s.  We plot the velocity, radiation and gas energy
densities and gas density versus position.  The vertical dashed lines 
show the expected shock thickness.  Gas flows into the shock from the left.
The after effects of the transient that occurs at the start of the
calculation can be seen at the far right of the plots.    }
\end{figure}
 
Figure \ref{fig:radthick} shows the results at a time $t= 5 \times 10^{-4}$~s,
of the radiation energy density $E$, gas energy density $e$, velocity $v$, and
density $\rho$ versus position. The vertical dashed lines indicate the expected
shock thickness and the SPH results are in good agreement.   The after-effects
of the transient moving downstream can again be seen on the right of the plots
of density and gas energy density.  \citet{WB2004} were unable to run this test
case with the large timestep used here because it required an unacceptable
amount of computational time.  The calculation presented here required
approximately a week.

\section{Conclusions}

We have presented a more efficient method for performing radiative transfer in
the flux-limited diffusion approximation within the SPH formalism. 
This gives a speed increase of
many thousand times over the code of \citet{WB2004}.  In every test,
the new implicit code is much faster than the old code for large implicit 
timesteps, with no loss in accuracy.

Whilst the method described here is presented in one dimension, the addition of
the algorithm into a three-dimensional code is easily accomplished. The major
difference between the one- and three-dimensional algorithms is the form of the
radiation pressure, which involves a more complicated tensor equation. We are
performing simulations in three dimensions using this algorithm which will
be published in due course.

\section*{Acknowledgments}

SCW acknowledges support from a PPARC postgraduate studentship. 
MRB is grateful for the support of a Philip Leverhulme Prize.

\bibliographystyle{mn2e}
\bibliography{MF741rv}

\end{document}